\documentclass[12pt]{article}
\usepackage{graphicx}

\newcommand\pubnumber{WSU--HEP--XXYY}
\newcommand\pubdate{\today}

\def\wayne{Experimental Physics Division\\
Institute of High Energy Physics, Beijing, 100049, China}
\def\support{\footnote{Work supported in part
by the CAS/SAFEA International Partnership Program for Creative
Research Teams, CAS and IHEP grants for the Thousand/Hundred Talent programs and National Natural Science Foundation of China under Contracts Nos. 11175189 and 11125525.}}

\def\Title#1{\begin{center} {\Large #1 } \end{center}}
\def\Author#1{\begin{center}{ \sc #1} \end{center}}
\def\Address#1{\begin{center}{ \it #1} \end{center}}

\newcommand\pubblock{\rightline{\begin{tabular}{l} \pubnumber\\
         \pubdate  \end{tabular}}}
\newenvironment{Abstract}{\begin{quotation}  }{\end{quotation}}
\newenvironment{Presented}{\begin{quotation} \begin{center} 
             PRESENTED AT\end{center}\bigskip 
      \begin{center}\begin{large}}{\end{large}\end{center} \end{quotation}}
\def\Acknowledgements{\bigskip  \bigskip \begin{center} \begin{large}
             \bf ACKNOWLEDGEMENTS \end{large}\end{center}}

\def\beq{\begin{equation}}
\def\eeq#1{\label{#1}\end{equation}}
\def\eeqn{\end{equation}}

\def\beqa{\begin{eqnarray}}
\def\eeqa#1{\label{#1}\end{eqnarray}}
\def\eeqan{\end{eqnarray}}

\let\bar=\overbar

\def\Dslash{\not{\hbox{\kern-4pt $D$}}}
\def\dslash{\not{\hbox{\kern-2pt $\del$}}}

\def\msb{{\bar{\ssstyle M \kern -1pt S}}}

\begin{document}
\begin{titlepage}
\pubblock

\vfill
\Title{Search for low-mass Higgs and dark photons at BESIII}
\vfill
\Author{Vindhyawasini Prasad\support, Haibo Li and Xinchou Lou }
\Address{\wayne}
\vfill
\begin{Abstract}
Many extensions of the Standard Model introduce a new type of weak-interacting degrees of freedom. These models are motivated by the results of recent experimental anomalies. Typical models, such as Next-to-Minimal Supersymmetric Standard Model and Light Hidden Dark-sector Model, introduce the possibilities of low-mass Higgs and dark bosons. The masses of such particles are expected to be few GeV and thus making them accessible at BESIII experiment, an $e^+e^-$ collider experiment running at tau-charm region. This report summarizes the recent results of  low-mass Higgs and dark bosons searches at BESIII.  

\end{Abstract}
\vfill
\begin{Presented}
The 7th International Workshop on Charm Physics (CHARM 2015)\\
Detroit, MI, 18-22 May, 2015
\end{Presented}
\vfill
\end{titlepage}
\def\thefootnote{\fnsymbol{footnote}}
\setcounter{footnote}{0}
%

\section{Introduction}
Many extensions of the Standard Model (SM), such as light hidden dark-sector model \cite{hidden} and Next-to-Minimal Supersymmetric Standard Model (NMSSM) \cite{nmssm}, introduce light weak-interacting degrees of freedom. In the hidden dark-sector model, WIMP-like fermionic dark matter particles are charged under a new force carrier. The corresponding gauge field, the dark photon ($\gamma'$), is coupled to the SM particles via a kinetic mixing \cite{epsilon} with mixing strength ($\epsilon$). In the framework of this hidden sector, dark matter particles annihilate into a pair of dark photons, which subsequently decay of SM particles. The mass of the dark photon is constrained to be at most a few GeV to accommodate the features of recent experimental anomalies observed in cosmic rays \cite{pamela}. 

The NMSSM adds an additional singlet super field to the Minimal Supersymmetric Standard Model (MSSM) to solve the so called naturalness problem of the MSSM \cite{MSSM}. The Higgs sector of the NMSSM contains three CP-even Higgs bosons, two CP-odd Higgs bosons and two charged Higgs boson. The mass of the lightest CP-odd Higgs boson ($A^0$) can be less than twice the mass of a charm quark and produced via $J/\psi \rightarrow \gamma A^0$ \cite{wilzeck}. The coupling of the Higgs field to down-type (up-type) quark pair is proportional to tan$\beta$ (cot$\beta$), where tan$\beta$ is the ratio of up and down type of Higgs doublets. The branching fraction of $J/\psi \rightarrow \gamma A^0$  decay is expected to be in the range of $10^{-9} - 10^{-7}$, depending on $A^0$ mass and coupling \cite{bra0}. The final state to which the $A^0$ decays depends on various parameters such as tan$\beta$ and $A^0$ mass \cite{Dermisek}. 

The discovery of such a low-mass states would open a new frontier in particle physics beyond the SM. BABAR \cite{babardarkphot,babarlighthigh09,babarlighthiggs13}, Belle \cite{belledarkphot}, CMS \cite{cmslighthiggs} and CLEO \cite{cleohiggs} experiments have performed the search for these low-mass states and reported negative results so far. More recently, BESIII has also performed the search for low-mass Higgs and dark bosons using the  data collected at $J/\psi$, $\psi(2S)$ and $\psi(3770)$ resonances. The following sections describe the detail information about these new physics searches at BESIII.  

\section{Search for a low-mass dark photon}
The high intensity $e^+e^-$ collider experiments can provide a clean environment to probe the low-mass dark-sectors \cite{Batell, Essig}. The width of the dark photon is suppressed by a factor of $\epsilon^2$ and expected to be very narrow than the experimental resolution. The dark photon can therefore be detected via the initial state radiation (ISR) process of $e^+e^- \rightarrow \gamma l^+l^-$ ($l = e, \mu$). BABAR has performed the search for a dark photon using this ISR decay process and set one of best exclusion limits in the mass range of $0.02 - 10.2$ GeV/$c^2$ \cite{babardarkphot}. Due to collecting a large amount of data sample at different center-of-mass energies, BESIII has the capability to produce the compatible exclusion limits on the coupling of the dark photon to the SM particles in the absence of any signal observations. 

The search for di-lepton decays of a light dark photon is performed in the ISR process of $e^+e^- \rightarrow \gamma_{ISR} \gamma' \rightarrow l^+l^-$ ($l=e, \mu$) using $2.9$ fb$^{-1}$ of $\psi(3770)$ data collected by BESIII experiment during $2010-2011$. The events of interest are selected with two oppositely charged tracks only, where both the tracks are  identified as a muon (electron) using the standard muon (electron) particle identification (PID) system developed by the BESIII experiment \cite{muonpid}. In order to remove the beam related backgrounds, both the tracks are required to be originated within 10.0 cm along the beam direction and 1.0 cm in the transverse direction of the beam. The polar angle of  both the tracks is required to be in the main drift chamber (MDC) acceptance region (i.e. $|cos\theta| < 0.93$). 

\begin{figure}[htb]
\centering
\includegraphics[height=2.0in,width=2.9in]{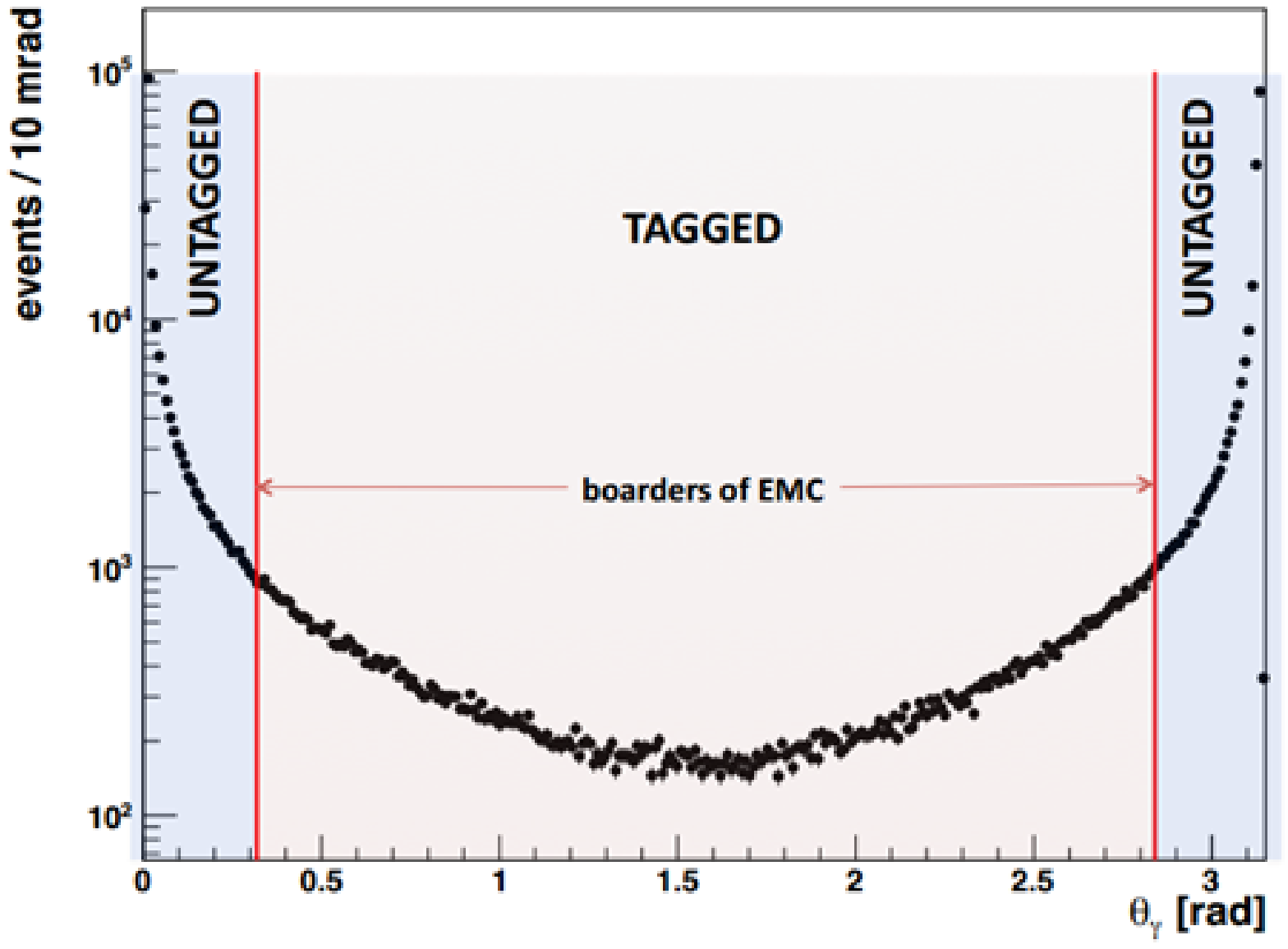}
\includegraphics[height=2.0in,width=2.9in]{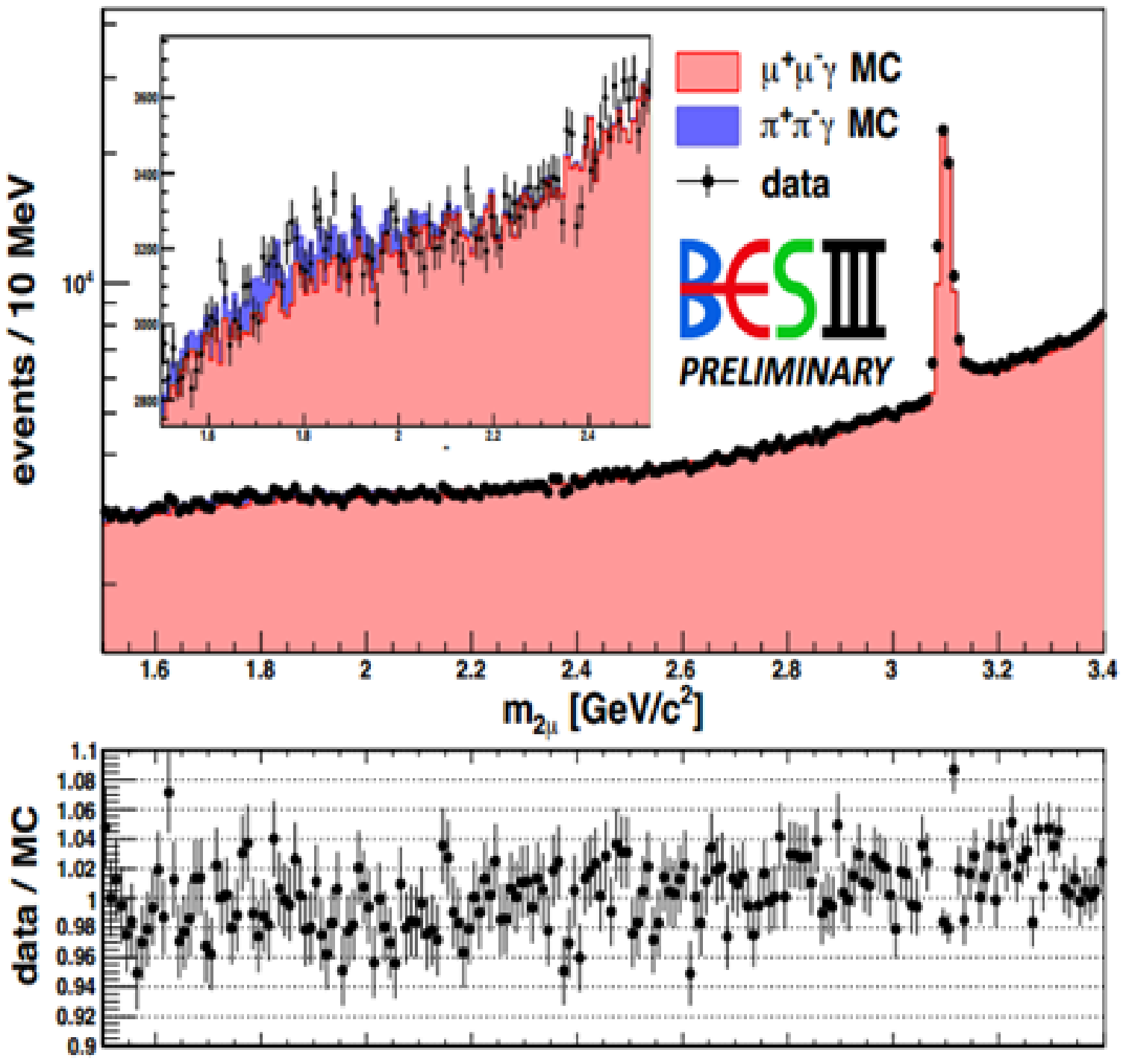}
\caption{(Left) Polar angle ($\theta_{\gamma}$) distribution of the ISR photon and (right) the di-muon invariant mass ($m_{2\mu}$) distribution of data (black dot points) and MC (shaded area).  The marked area around the $J/\psi$ resonance is excluded from the analysis. The inlay in the upper left in the right hand side plot displays an enlargement of the $m_{2\mu}$ distribution. The region of the $\theta_{\gamma}$ distribution in the BESIII detection region is $[0.376,2.765]$ radian. In the untagged photon method, the  ISR photon is required to be not detected in the EMC detection region.  }
\label{fig:ISRGam}
\end{figure}

This search is based on an untagged ISR photon method,  where the ISR photon is emitted at a small polar angle of photon and not detected within the angular acceptance of the  electromagnetic  calorimeter (EMC), to suppress the non-ISR backgrounds (Figure~\ref{fig:ISRGam} (left)). In order to improve the mass resolution of the dark photon, a one-constraint (1C) kinematic fit is performed with two charged tracks and missing track with a condition that the mass of the mission track must be zero. The fit quality condition $\chi_{1C}^2 < 20$ ($\chi_{1C}^2 < 5$)  is applied in the $\gamma_{ISR} \mu^+\mu^-$ ($\gamma_{ISR}e^+e^-$) case. We finally require that the di-lepton invariant mass must be within $1.5-3.4$ GeV/$c^2$. Figure~\ref{fig:ISRGam} (right) shows the plot of di-muon invariant mass distribution of both data and Monte Carlo (MC) samples. No evidence of dark photon production is found in any of the decay processes at any mass points and we set $90\%$ C.L. upper limits on $\epsilon$ as a function of $m_{\gamma'}$ using the formula of Equation 19 in \cite{epsformula}. 

\section{Search for a low-mass Higgs boson}
Previous searches of the $A^0$ performed by BABAR \cite{babarlighthigh09,babarlighthiggs13}, CLEO \cite{cleohiggs},  and CMS \cite{cmslighthiggs} experiments have placed very strong exclusion limits on the coupling of the b-quark to the $A^0$ \cite{Dermisek,cmslighthiggs,babarlighthiggs13}. BESIII has also previously searched for  di-muon decays of a light Higgs boson in the radiative decays of $J/\psi$ using the data at $\psi(2S)$ resonance, where the $J/\psi$ events were selected by tagging the pion pair from $\psi(2S) \rightarrow \pi^+\pi^- J/\psi$ transitions \cite{BESIII_Higgs0}. No evidence of $A^0$ candidate was found and exclusion limits on
   $\mathcal{B}(J/\psi \rightarrow \gamma A^0) \times \mathcal{B}(A^0 \rightarrow \mu^+\mu^-)$ were is set in the range of $(0.4-21.0) \times 10^{-6}$ for $0.212 \le m_{A^0} \le 3.0$ GeV/$c^2$, which seem to be above the theoretical prediction. The large data samples collected at the $J/\psi$ resonance by BESIII experiment allow us to perform this study once again with a high precision.

 The search for a light Higgs boson is performed in the fully reconstructed decay process of $J/\psi \rightarrow \gamma A^0$, $A^0 \rightarrow \mu^+\mu^-$ using 225 million $J/\psi$ events collected by BESIII experiment. Same amount of generic $J/\psi$ decays of simulated MC events is used for the background studies. This work is based upon a blind analysis; we don't see the full data sample unless all the selection criteria are finalized. The event of interests is selected with two oppositely charged tracks and at least one good photon candidate. The minimum energy of this photon is required to be 25 MeV in the barrel region ($|cos\theta| < 0.8$) and 80 MeV in the end-cap region ($0.86 < |cos\theta| < 0.92$).   The EMC time is also required to be in the range of $[0,14](\times 50)$ ns within the event start time to suppress the electronic noise and energy deposits unrelated to the signal events.  Additional low energetic photons are allowed to be in the events. In order to reduce the beam related backgrounds, charged tracks are required to originate within $\pm 10.0$ cm from the interaction point in the beam direction and within $\pm 1.0$ cm in perpendicular to the beam. The charged tracks are required to be in the polar angle region, $|cos\theta| < 0.93$, thus having a reliable measurement in the MDC. 

The two charged tracks are assumed to be the muons and required at least one charged track must be identified as muon using a muon PID system, which is based on the selection criteria of following three variables: (1) the energy deposited in the EMC ($E_{cal}^{\mu}$) by a muon particle must be within the range of  $0.1 < E_{cal}^{\mu} < 0.3$ GeV, (2) the absolute value of the time difference between time-of-flight (TOF) and expected muon time ($\Delta t^{TOF}$) must be less than 0.26 ns and (3) the penetration depth in muon counter (MuC) must be greater than $(-40.0 +70\times p$) cm for  $0.5 \le p \le 1.1$ GeV/$c$ and  40 cm for $p > 1.1$ GeV/$c$.  A 4C kinematic fit is performed with two charged tracks and one of the good photon candidates to improve the mass resolution of the $A^0$ candidate. The $\chi^2$ from the 4C kinematic fit is required to be less than $40$ to suppress the background contributions from the decay processes of $J/\psi \rightarrow \rho\pi$ and $e^+e^- \rightarrow \gamma \pi^+\pi^-\pi^0$. We further require that either one of the tracks must have cosine of muon helicity angle (cos$\theta_{\mu}^{hel}$), defined as the angle between the direction of one of the muons and the direction of $J/\psi$ in the $A^0$ rest frame, to be less than 0.92 to suppress the backgrounds peaking in the forward direction.   

\begin{figure}[htb]
\centering
\includegraphics[height=2.0in,width=3.2in]{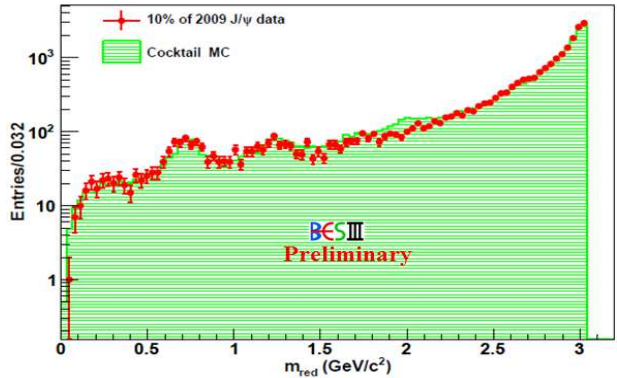}
\caption{The $m_{red}$ distribution of cocktail MC (shaded area) and $10\%$ of the $J/\psi$ data (red error bars). The cocktail MC sample is a composition of the simulated events of generic $J/\psi$ decays and $e^+e^- \rightarrow \gamma \mu^+\mu^-$. The MC samples are normalized with data luminosity. Due to very low statistic in the $10\%$ of the $J/\psi$ data, an additional source of the peaking background at the $f_4(2050)$ mass position seems to be disappeared. However, this background might be seen in the full $J/\psi$ data sample.}
\label{fig:mred}
\end{figure}

Figure~\ref{fig:mred} shows the reduced di-muon mass, $m_{\rm red} = \sqrt{m_{\mu^+\mu^-}^2 -4m_{\mu}^2}$, distribution of the $10\%$ of $J/\psi$ data together with a composite MC sample of generic $J/\psi$ decays and ISR production of $e^+e^- \rightarrow \gamma \mu^+\mu^-$ simulated events. The $m_{\rm red}$ is equal to twice the muon momentum in the $A^0$ rest frame, easier to model it near the threshold in comparison to the di-muon invariant mass. In  $10\%$ of the $J/\psi$ data-set, the background is dominated by following two types of process: \rm{\lq\lq non-peaking\rq\rq} component of $e^+e^- \rightarrow \gamma \mu^+\mu^-$ and \rm{\lq\lq peaking\rq\rq}  components of $J/\psi \rightarrow \rho\pi$ and $J/\psi \rightarrow \gamma X$ decays, where $X = f_2(1270)$ and $f_0(1710)$ mesons. However, an addition peaking component at $f_4(2050)$ mass position seems to be appearing in the MC sample of generic $J/\psi$ decays. The additional source of this background can be seen in the full $J/\psi$ data-set, we therefore also consider this source of background to develop our maximum likelihood (ML) fitting procedures to extract the signal events as a function of $m_{A^0}$. 

We use the sum of the two Crystal Ball (CB) to model the $m_{red}$ distribution of the signal while generating the signal MC samples at $23$ assumed $A^0$ mass points. The $m_{red}$ resolution of the signal varies in the range of $2-12$ MeV/$c^2$ while the signal efficiency varies from $49\%$ to $33\%$ depending upon the momentum values of two muons at different Higgs mass points. The $m_{red}$ distribution of the non-peaking background is modeled by the $n^{th}$ ($n=2,3,4,5$) order Chebyshev polynomial function depending upon different $m_{A^0}$ intervals. In the low-mass region, we constrain the curve must pass through the origin when the cross-section is zero (i.e. $m_{red} \approx 0$). We $m_{red}$ distribution of the $\rho$ resonance is modeled by a Cruijff function \cite{cruijff} and  $f_2(1270)$, $f_0(1710)$ and $f_4(2050)$ resonances by the sum of the two CB functions. 

The additive and multiplicative types of the systematic uncertainties are considered in this analysis.  An additive uncertainty arises due to fit bias and fixed PDF parameters, and does not scale with the number of reconstructed signal events. The multiplicative uncertainties scale with the number of reconstructed signal events and arise due to the reconstruction efficiency, the uncertainty in the number of $J/\psi$ mesons ($1.30\%$), the resolutions of the peaking backgrounds ($1.2\%$ for the $\rho$ resonance and $6.7\%$ for $f_2(1270)$, $f_0(1710)$ and $f_4(2050)$ resonances) and the PDF parameters of the non-peaking backgrounds ($3.18\%$). 

The uncertainty associated with the photon detection efficiency is measured to be  less than  $1\%$ using a control sample of the ISR process of $e^+e^- \rightarrow \gamma \mu^+\mu^-$, where the ISR photon is estimated using the four-momenta of two charged tracks \cite{photon_eff}. We use a control sample of $J/\psi \rightarrow \mu^+\mu^- (\gamma)$, in which one track is tagged with tight muon PID, to study the systematic uncertainty associated with the muon PID($(4.0-5.73)\%$), $\chi_{4C}^2$ ($1.56\%$) and the cos$\theta_{\mu}^{heli}$ ($0.34\%$) requirements. The final value of muon PID uncertainty also takes into account the fraction of events with one track or two tracks identified as muons, which is obtained from the signal MC. The total additive and multiplicative uncertainties vary in the range of  ($0.502 - 0.767$) events and $(5.95 - 8.96)\%$, respectively depending on $m_{A^0}$.     

\begin{figure}[htb]
\centering
\includegraphics[height=2.0in,width=2.9in]{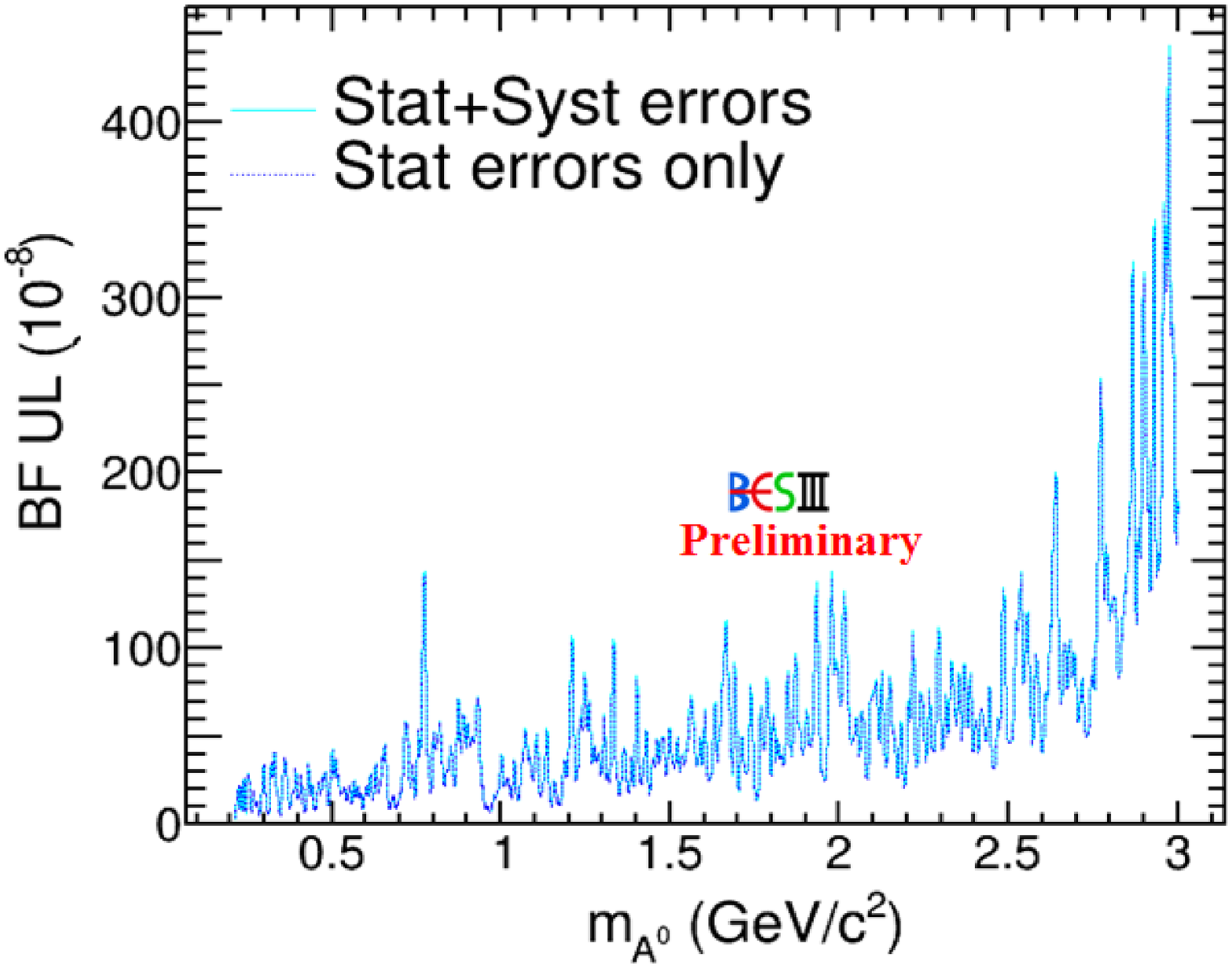}
\includegraphics[height=2.0in,width=2.9in]{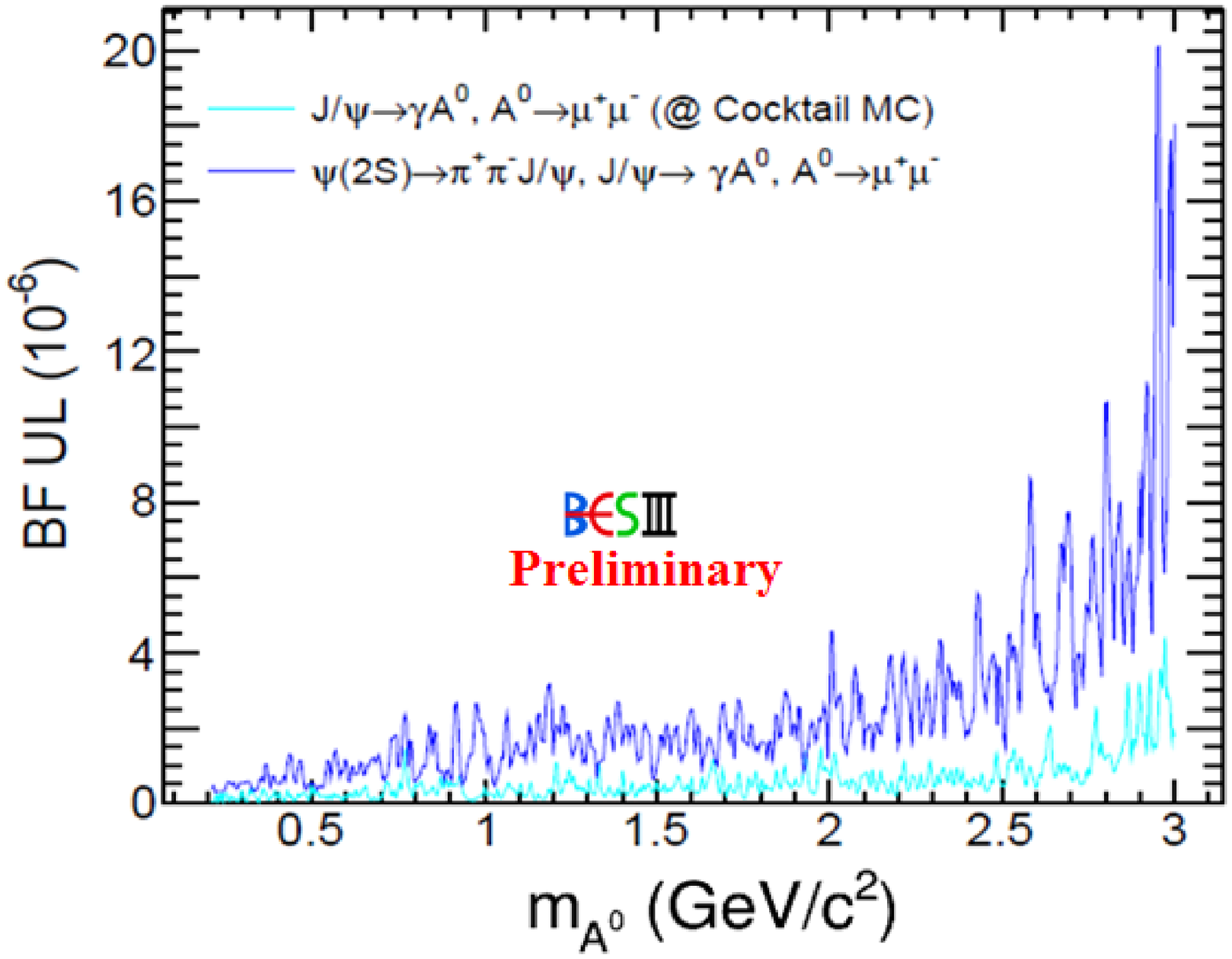}
\caption{The $90\%$ C.L. upper limits of the product branching fraction $\mathcal{B}(J/\psi \rightarrow \gamma A^0) \rightarrow (A^0 \rightarrow \mu^+\mu^-)$ as a function of $m_{A^0}$ for new BESIII measurement, including with and without systematic uncertainty  shown by cyan and blue colors, respectively (left) and old  versus new BESIII measurement (right). The new BESIII exclusion limits are based on a cocktail MC sample containing the same statistics as we expect the full data sample.   }
\label{exclusion:lim}
\end{figure}

We perform the search for a narrow resonance in the steps of  half of $m_{red}$ resolution at 2035 mass points using the cocktail MC sample. The systematic uncertainty is included with the final results by convolving the negative log likelihood (NLL) versus branching fraction curve with a Gaussian distribution having a width equal to the systematic uncertainty. The projected $90\%$ C.L. upper limits on the product branching fraction $\mathcal{B}(J/\psi \rightarrow \gamma A^0 \times \mathcal{B}(A^0 \rightarrow \mu^+\mu^-)$ are observed to be in the range of $(2.8 - 386.5)\times 10^{-8}$ for $0.212 \le m_{A^0} \le 3.0$ GeV/$c^2$ depending upon the $m_{A^0}$ mass points (Figure~\ref{exclusion:lim} (left)).  The new BESIII expected limits seem to have an order of magnitude improvement than the previous measurement  \cite{BESIII_Higgs0} (Figure~\ref{exclusion:lim} (right)) and can significantly constrain the parameters of the new physics models \cite{bra0,new_physics}. We will unblind the full $J/\psi$ data sample very soon to produce the final results.

\section{Summary and conclusion}
The BESIII has performed the search for di-lepton decays of the low-mass Higgs and dark bosons using the data samples collected at $J/\psi$ and $\psi(3770)$ resonances, respectively. No evidence of any narrow resonance particles is found in the data-sets and set one of the stringent exclusion limits. These exclusion limits can constrain a large fraction of the parameters in the new physics models including the NMSSM. All the results are preliminary. BESIII is performing  the search for new physics in other decay processes too, and we are looking forward to seeing some more exciting results in the near future.   

\Acknowledgements
This work is supported in part by the CAS/SAFEA International Partnership Program for Creative Research Teams, CAS and IHEP grants for the Thousand/Hundred Talent programs and National Natural Science Foundation of China under the Contracts Nos. 11175189 and 11125525.


\end{document}